\batchmode
\makeatletter
\def\input@path{{/home/simon/Documents/Papers/BayesAdapt//}}
\makeatother
\documentclass[english]{article}
\usepackage[T1]{fontenc}
\usepackage[latin9]{inputenc}
\usepackage{geometry}
\usepackage{float}
\usepackage{graphicx}
\usepackage{setspace}
\usepackage{amssymb}
\usepackage[authoryear]{natbib}

\makeatletter

\floatstyle{ruled}
\newfloat{algorithm}{tbp}{loa}
\floatname{algorithm}{Algorithm}

\usepackage{babel}
\makeatother

\begin{document}

\title{A flexible Bayesian method for adaptive measurement in psychophysics}

\author{Simon Barthelmé, Pascal Mamassian}

\maketitle
\begin{abstract}
In psychophysical experiments time and the limited goodwill of participants
is usually a major constraint. This has been the main motivation behind
the early development of adaptive methods for the measurements of
psychometric thresholds. More recently methods have been developed
to measure whole psychometric functions in an adaptive way. Here we
describe a Bayesian method to measure adaptively any aspect of a psychophysical
function, taking inspiration from Kontsevich and Tyler's \citeyearpar{KT1999}
optimal Bayesian measurement method. Our method is implemented in
a complete and easy-to-use MATLAB package. 
\end{abstract}

\section{Motivation}

Over the years a large number of methods have been developed with
the goal to obtain the most information about an observer's behaviour
in a limited amount of time. They range from simple empirical schemes
such as the staircase, to more elaborate mechanisms with better theoretical
foundations, such as the QUEST or ZEST methods \citep{QUEST,ZEST}.
Some only estimate the threshold, while others aim to provide information
about the slope of the psychometric function as well \citep{KT1999,KujalaLukka06}.

Here we build on the latter to provide a very general framework for
adaptive measurement and fitting of psychometric functions. We provide
the option of focusing on a just one aspect of a psychometric function
(for example, threshold or slope alone), or on the whole psychometric
function. By using the framework of Bayesian inference, we provide
theoretically well-founded methods for estimating the parameters and
obtaining maximal information about them. An additional concern has
been with the development of fast algorithms so that waiting time
between trials could be minimised. We provide an easy-to-use software
package in MATLAB for researchers without the time, the required technical
background, or the inclination to implement our algorithms.

\section{Description of the method}

\subsection{General framework}

\begin{figure}
\includegraphics[scale=0.7]{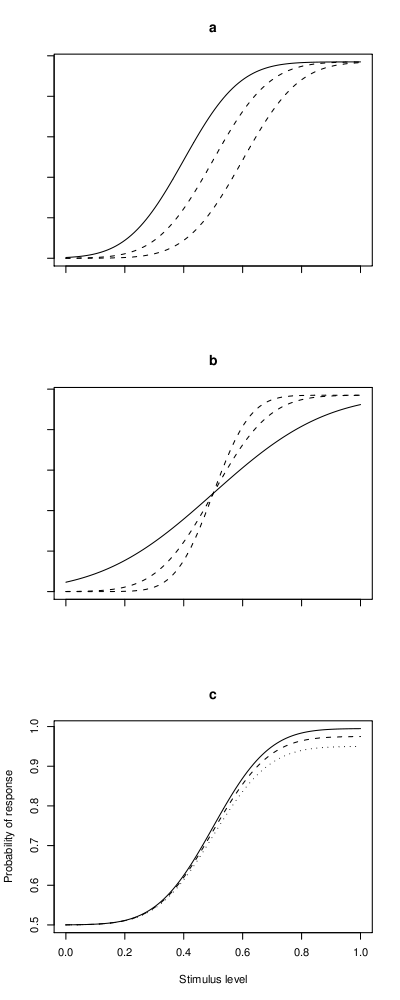}\caption{Effects of varying the parameters of the psychometric function. a.
Increasing location, $\mu$ b. Increasing inverse slope $\sigma$
c. Increasing lapse rate $\lambda$ \label{fig:parameters-psi}}

\end{figure}

A psychometric function links the stimulus level (for example, contrast
or signal-to-noise ratio) to a probability of response (often, the
probability of a correct response). It can be defined by three parameters:
one for the position along the stimulus level axis, one for the slope,
and one for the lapse rate. The effects of the three are illustrated
on figure \ref{fig:parameters-psi}. Thresholds are defined in the
following way: first pick a performance level, for example 75\%. Then
the corresponding threshold is given by the inverse of the psychometric
function, ie, it is the stimulus level such that the probability of
response is 0.75.

The aim of a acquiring responses from an observer is to estimate some
feature of the psychometric function, either its three parameters,
or just its threshold or slope. The non-adaptive way to do this is
for the experimenter is to spend some time in a dark room playing
around with the stimulus and then choosing within a range a number
of stimulus values they think appropriate to measure observer's performance.
This is known as the method of constant stimuli. The usual result
is that due to variability between observers, and due also to the
fact that the experimenter, having spent countless hours doing the
task, will overestimate everybody else's performance, thinking that
1\% contrast should be easy enough for anyone. The method of constant
stimuli thus leads to very wasteful data collection, with observers
being tested at chance performance, or close to 100\%. In the worst-case
scenario, none of the datapoints are actually informative. 

Despite this flaw, the method of constant stimuli enjoys the considerable
resilience of a gold standard, notably because it results in fits
that are easy to evaluate visually - but this is sometimes misleading,
as we explain in section \ref{sub:I-can't-check-the-fits} below.
The general class of alternatives to the method of constant stimuli
is known in psychophysics as \emph{adaptive methods},\emph{ }and in
other circles as \emph{active learning} \citep{Cohn95:ActiveLearning}.
Adaptive methods take advantage of the intuitive idea that if an observer
has already answered 10 out of 10 times correctly at a given level,
then there isn't much point in further interrogation.

\begin{figure}
\includegraphics[clip,scale=0.7]{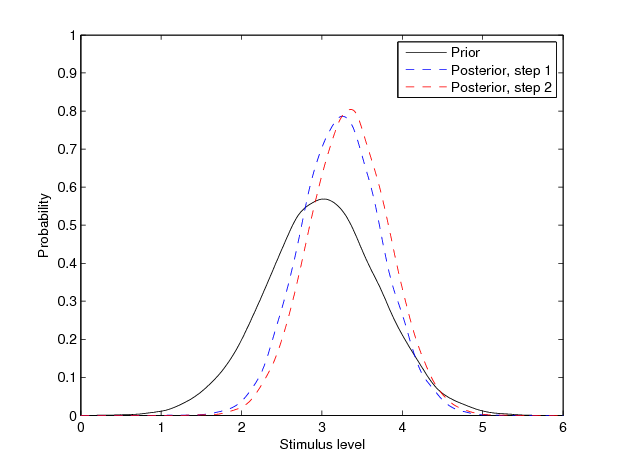}

\caption{\label{fig:illus-post-threshold}An illustration of Bayesian belief
updating. We start out with a prior for the 75\% threshold centered
around 3 (black solid curve). On the first trial, we test the observer
at stimulus level 6 and get an incorrect response. We update our beliefs
according to Bayes' theorem, yielding the first posterior distribution
(blue dashed curve). The distribution agrees with intuitive reasoning:
an incorrect response at level 6 would be unlikely if the real threshold
were below 3, but more likely if it were above. On the 2d trial, the
posterior from our first trial becomes our prior, and we measure a
correct response at level 3, which is relatively likely under most
threshold values between 2 and 4, and the posterior probability (red
dashed curve) is changed only slightly. Notice how the uncertainty
decreases over time, first fast, then more slowly. This is a general
feature of Bayesian updating.}

\end{figure}

\begin{figure}
\includegraphics[clip,scale=0.7]{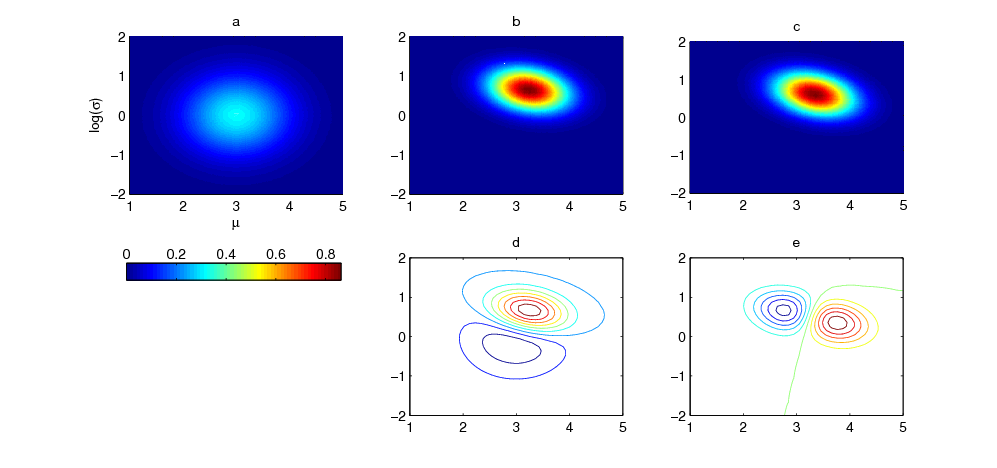}\caption{\label{fig:illus-post-mu-nu}For the same data illustrated in figure
\ref{fig:illus-post-threshold}, we show the Bayesian update in $\mu,\log\sigma$
space (see \ref{sub:Functional-form}). a. Marginal prior over $\mu,\log\sigma$
b. Posterior after first trial c. Posterior after 2d trial. d. Difference
between prior and first posterior e. Difference between second and
first posterior. The updates are obtained via the Laplace approximation.}

\end{figure}

In Bayesian adaptive methods, the measurement process is viewed as
the step-by-step update of a probability distribution over the object
of interest. Imagine we want to measure the 75\% threshold: we start
out by expressing our prior information about potential thresholds
as a prior distribution. Just as in the case of the method of constant
stimuli, we use the information acquired in the hours spent fiddling
around and exploiting next-door colleagues to get some idea of what
would be a range of realistic thresholds for other observers. This
information will be udpated every time we collect a response from
the observer: we update our beliefs according to whatever data we
get. How exactly we need to update our beliefs is given by Bayes'
theorem. The process is illustrated in figure \ref{fig:illus-post-threshold}
and \ref{fig:illus-post-mu-nu}.

The uncertainty that remains about the threshold (or about any other
set of parameters) is given by the entropy of the distribution. A
distribution that has a single peak - indicating that we are extremely
sure about the value of our parameter - will have minimum entropy,
while a flat distribution - indicating complete lack of confidence
- has maximum entropy. A decrease in entropy, then, is equivalent
to an increase in confidence. At each step we ask: what is the stimulus
level so that the confidence will increase maximally on average? Obviously
if we know that the probability that the observer's threshold is over
20\% is 0.01, then there is no point in picking a stimulus level in
the vicinity of 20\%. The right stimulus level is given by a minimum
in a cost function, which can be computed by the algorithms we detail
below.

\subsection{Functional form for the psychometric function\label{sub:Functional-form}}

For psychometric functions in forced-choice experiments we use the
following form%
\footnote{In Yes/No experiments, a slightly different form must be used, see
Appendix%
}:

\begin{equation}
\Psi(x;\mu,\sigma,\lambda)=(1-\lambda)\left(\gamma+(1-\gamma)\Phi(x;\mu,\sigma)\right)+\lambda\gamma\label{eq:psychfuncFC}\end{equation}

\begin{itemize}
\item \emph{x} is the stimulus level
\item $\mu$ and $\sigma$ describe the placement and slope of the psychometric
function
\item $\gamma$, a constant, is the chance rate. In 2AFC designs, $\gamma$
is set to 0.5.
\item $\Phi(x;\mu,\sigma)$ is the cumulated Gaussian function. This choice
of shape is on aesthetic grounds, choosing for example a logistic
shape makes little to no difference in practice. 
\item Following the recommendations in \citet{Wichmann2001B,Wichmann2001A}
we include a lapse term, $\lambda$. Here we assume that on every
trial, the observer either lapses and makes a random choice (with
probability of success $\gamma$) or behaves in a stimulus-dependent
manner, with probability of success $\gamma+(1-\gamma)\Phi(x;\mu,\sigma)$.
Including a lapse term is useful in order to insure robustness against
random errors at high stimulus levels.
\end{itemize}
One can measure psychometric functions with different goals in mind.
If the psychometric function is the independent variable in an experiment,
one will wish to know its parameters $\theta=(\mu,\sigma,\lambda)$
as well as possible, using as few datapoints as possible. This is
the aim of Kontsevich and Tyler's original method. In another scenario,
one wishes only to measure thresholds, for example to set the stimulus
level in an experiment so that the difficulty is more or less the
same for all observers, or simply because the threshold provides a
useful summary of performance. Sometimes one may want to know more
than just one threshold, for example both the $.73$ and the $.65$
thresholds could be of interest: we show in section \ref{sub:Reduction-mult-thresh}
that this reduces to the $\Psi$ method solution.

\subsection{The $\psi$ method }

\subsubsection{Description}

The $\Psi$ method we describe here was first developed by \citet{KT1999}
and improved on by \citet{KujalaLukka06}. We frame the problem in
Bayesian terms: our knowledge of the parameters of the psychometric
function is described by a probability distribution over the parameters
$\theta$. We start out with prior beliefs about $\theta$, descriped
by a prior pdf $p_{0}(\theta)$, and at each step we collect data
$y_{t}=(x_{t,}r_{t})$ where $r_{t}$ is the response and $x_{t}$
the stimulus level. We then update our beliefs according to Bayes'
theorem. We define $\mathbf{y_{t}}=(y_{1},y_{2},...,y_{t})$

\[
p_{t}(\theta|\mathbf{y_{t}})=\frac{p(r_{t}|\theta)p_{t-1}\left(\theta|\mathbf{y_{t-1}}\right)}{p(\mathbf{y_{t})}}\]

From now on when not required for clarity we will drop the dependency
of $\theta$ on $\mathbf{y}$ and simply write $p_{t}(\theta)$ for
our posterior over $\theta$ at time $t$.

At any point in time, our uncertainty over the real value of the parameters
can be summarised by the entropy of the posterior distribution:

\[
H_{t}(\theta)=-\int p_{t}(\theta)\ln p_{t}(\theta)d\theta\]

The goal of the adaptive procedure must be to find the optimal stimulus
placement so that the entropy is as low as possible when we are done
collecting data. The optimal way of doing that would be to find at
each time step the stimulus level that minimises expected entropy
over all the remaining steps \citep{Pelli87Quest}. This is computationally
intractable and Nontsevick and Tyler's $\Psi$ method only minimises
the expected entropy over the next step, ie. it finds a $x_{t+1}$such
that:

\[
E_{r_{t+1}}\left(H_{t+1}(\theta)\right)\]

is minimised. Here $E_{r_{t+1}}$denotes the expectation over the
unknown response $r_{t+1}$, given a stimulus level of $x_{t+1}$.

Intuitively, the $\Psi$method finds a stimulus level such that the
observer's response will have maximum effect (in expectation) on our
beliefs. For example, if we know from our posterior on $\theta$ that
the observer will be very close to 100\% correct at a given stimulus
level, then there is little point in placing the next trial at that
level because the response is unlikely to teach us something new.

\subsubsection{Computational issues}

\paragraph{Computation of the posterior distribution}

A number of computational issues arise in applying the $\Psi$ method.
First, there is no closed form for the posterior $p_{t}(\theta)$,
whatever the choice of prior, which makes it difficult to compute.
Kontsevich \& Tyler choose to discretise the parameter space, which
leads to a difficult trade-off between precision and computation time
and to many problems when parameter values lie near the boundaries.
In their extension to the $\Psi$ method, Kujalla \& Lucca use particle
filtering to represent and update the posterior. Here we use a Laplace
approximation \citep{McKayITILG}, which is computationally very cheap
and quite accurate for the right choice of parameterisation (see below).
Given an unnormalised density $P(\theta)\propto p(\mathbf{y}|\theta)p(\theta)$,
we find its mode $\theta^{*}$and approximate $P$ with $Q(\theta)$
a Gaussian centred on $\theta^{*}$. The covariance matrix of the
Gaussian is given by the inverse of the Hessian of $\log Q$ at the
mode. The approximation is accurate if $P(\theta)$ is unimodal and
roughly symmetrical around its principal axes. 

To improve the approximation should the need be felt, we provide the
option of using importance resampling \citep{GelmanBDA}. In importance
resampling, $n$ samples $\{q\}$ from the approximating density $Q(\theta)$
are drawn. Each is given a weight:

\[
w_{i}=\frac{P(q_{i})}{nQ(q_{i})}\]

We then draw $k<n$ samples without replacement from $\{q_{i}\}$
with probability $w_{i}$. In general this leads to an improvement
in the approximation of the posterior density, because points that
have low probability under $P(\theta)$ are less likely to be resampled.

\paragraph{Computation of the cost function}

In Kontsevich and Tyler's \citeyearpar{KT1999} original formulation,
the cost function for stimulus level is computed using algorithm \ref{cap:Computing-the-K&T-cf}.
The cost function is computed over a predefined, discrete set of stimulus
values and the best stimulus value is chosen. The algorithm is rather
slow because it must compute a posterior distribution twice for each
stimulus value. Kujalla \& Lucca give a way of using a Fourier transform
to reduce computational complexity. 

\begin{algorithm}
\caption{\label{cap:Computing-the-K&T-cf}Computing the K\&T cost function,
original algorithm}

For each stimulus level $x_{t+1}$

\begin{enumerate}
\item Compute the probability of a correct response: $p_{t}(r_{t+1}=1|x_{t+1})=\sum_{\theta}p_{t}(\theta_{i})\Psi(x;\theta)$
\item For $r_{t+1}=$ 0 and 1: compute a posterior $p(\theta|r_{t+1},x_{t+1},\mathbf{y})$
and its entropy $H(r_{t+1}|x_{t+1})$
\item Compute $c(x_{t+1})=p_{t}(r_{t+1}=1|x_{t+1})H(r_{t+1}=1|x_{t+1})+p_{t}(r_{t+1}=0|x_{t+1})H(r_{t+1}=0|x_{t+1})$
\end{enumerate}

\end{algorithm}

More usefully to us, they have devised a reformulation of the cost
function that enables fast computation when samples from the posterior
are available. At each time step, $\theta$ is a continuous random
variable with density given by $p_{t}(\theta)$. $r_{t+1}$, the observer's
next response, is a Bernoulli variable, whose probability of success
$\pi_{t}(x_{t+1})$depends on the stimulus level. Minimising the $\Psi$
cost function is equivalent to finding the $x_{t+1}$ that maximises
the mutual information between the response $r_{t+1}$ and $\theta$.
The mutual information is symmetrical, thus:

\begin{equation}
I(r_{t+1},\theta)=H(\theta)-H(\theta|r_{t+1})=H(x_{t+1})-H(r_{t+1}|\theta)\label{eq:MIresptheta}\end{equation}

The latter expression can be expanded as: 

\[
h\left(\pi_{t}(r_{t+1})\right)-\int p_{t}(\theta)h\left(\Psi(x_{t+1};\theta)\right)d\theta=h\left(\int p_{t}(\theta)\Psi(x_{t+1};\theta)d\theta\right)-\int p_{t}(\theta)h(\Psi(x_{t+1};\theta)d\theta\]

$h(\pi)$ is the entropy function for a Bernoulli variable $h(\pi)=-\pi\ln\pi-(1-\pi)\ln(1-\pi)$.

Given \emph{n }samples $\{q_{i}\}$from the posterior or its approximation,
an estimate of the cost function $c_{\Psi}$ is given by:

\begin{equation}
c_{\Psi}(x_{t+1})\approx h\left(n^{-1}\sum_{i=0}^{n}\Psi(x_{t+1},q_{i})\right)+n^{-1}\sum_{i=0}^{n}h\left(\Psi(x_{t+1},q_{i})\right)\label{eq:cost_KT_est}\end{equation}

The cost function can be computed with cost $\mathcal{O}(n)$.

\subsubsection{Optimisation of the cost function}

\begin{figure}
\includegraphics[clip,scale=0.5]{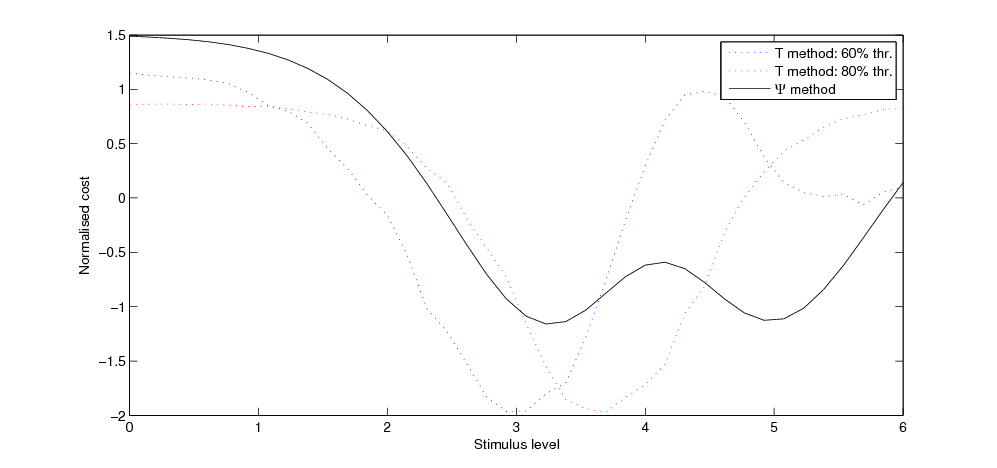}\caption{\label{fig:cost-functions}We simulate 100 trials of an adaptive run,
and plot the different cost functions giving the best placement for
the next trial according to different criteria. In the case of the
cost functions computed using the $T$ method, we plot the estimates
obtained using $6\times10^{4}$ samples. The noisy nature of the estimates
are still visible in the kinks of what would be otherwise smooth functions.
Only one function is convex over the range illustrated here, see text
for solutions to that problem.}

\end{figure}

The $\Psi$ cost function is in general not convex, and we have found
empirically that in some cases it exhibits two local minima, see figure
\ref{fig:cost-functions}. To find the global minima several strategies
can be used, including:

\begin{enumerate}
\item Sampling the function over a fixed grid, and starting a descent algorithm
from the best point. 
\item Using a descent algorithm with random restarts
\end{enumerate}
The whole process is summed up in algorithm \ref{algo:Psi-fast}.

\begin{algorithm}
\caption{\label{algo:Psi-fast}The $\Psi$ algorithm, fast version}

At each time step \emph{t}:

\begin{enumerate}
\item Find the maximum of the log posterior $p_{t}(\theta|\mathbf{y})$
\item Sample from the Laplace approximation to the log-posterior $p_{t}(\theta|\mathbf{y})$.
\item Find a minimum of the cost function $c_{\Psi}(x_{t+1})$, estimated
from the posterior samples and equation (\ref{eq:cost_KT_est}).
\item Obtain observer response at stimulus level $x_{t+1}$.
\end{enumerate}

\end{algorithm}

\subsection{Adaptive measurement for one attribute only: the $T$ method\label{sub:t-method}}

Often when measuring psychophysical performance only one attribute
is of interest: eg., the threshold, slope or range. The $T$ method
takes inspiration from methods such as ZEST \citep{ZEST} and the
$\Psi$ method described above, and provides an adaptive measurement
algorithm for such cases. We describe the algorithm in detail for
the case of measuring a threshold, the other two cases are similar.
It is possible to generalise the algorithm to handle any function
of the parameters into the real line, we show how below.

\subsubsection{One threshold}

\paragraph{A posteriori probability over a threshold\label{par:A-posteriori-probability-over-thresh}}

As we note above, quite often, the quantity of interest is the threshold,
defined as the stimulus level such that the probability of a correct
response is equal to $\alpha$. If the parameters $\theta^{*}$ are
known, then the threshold $\tau_{\alpha}$ is expressed as: \[
\tau_{\alpha}=\{x:\Psi(x;\theta^{*})=\alpha\}\]

For a given parameter vector $\theta^{*}$ there is only one $x$
that satisfies this equation: $\Psi(x;\theta^{*})$ has an inverse,
$\Psi^{-1}(\alpha;\theta^{*})$. 

However, we have an uncertainty over $\theta$ that translates into
an uncertainty over $\tau$. The posterior distribution over $\tau$
is given by:

\[
p(\tau_{\alpha}|\mathbf{y})=\intop_{\theta}\delta\left(\Psi(\tau_{\alpha};\theta)=\alpha\right)p(\theta|\mathbf{y})d\theta\]

This is difficult to compute directly because for a given $\tau$
there are many $\theta$ such that $\Psi(\tau_{\alpha};\theta)=\alpha$
- we would need to integrate along a curve in $\mathbb{R}^{3}$. Fortunately,
given samples from the posterior over $\theta$, it is easy to sample
from $p(\tau_{\alpha}|\mathbf{y})$. We simply run the samples through
$\Psi^{-1}$:

\[
\Psi^{-1}(\alpha;q)\sim p(\tau_{\alpha}|\mathbf{y})\,\textrm{if}\, q\sim p(\theta|\mathbf{y})\]

\paragraph{Cost function for one threshold\label{par:Cost-function-1-threshold}}

Just as in the case of the $\Psi$-method outlined above, we want
to maximise the quantity of information we gain at each step about
the threshold $\tau$. Here again we proceed by choosing the $x_{t+1}$that
maximises $I(r_{t+1},\tau)$, the mutual information between $r_{t+1}$
and the quantity of interest, $\tau$. There are a number of ways
to compute that mutual information, but a fast method is to sample
from the joint probability $p(r_{t+1},\tau)$, from the marginals
$p(r_{t+1})$ and $p(\tau)$ and to estimate the mutual information
from the samples. 

We explained above how to sample from $p(\tau)$. $p(r_{t+1})$ is
a Bernoulli random variable with probability of success $\int p_{t}(\theta)\Psi(x_{t+1};\theta)d\theta$.
Sampling from the joint distribution is achieved by getting $\{q_{i}\}$
samples from $p(\theta|y)$. Then a sample pair from the joint distribution
is obtained through: 

\[
\Psi^{-1}(\alpha;q_{i})\]

for the threshold and by simulating a Bernoulli variable with probability
of success $\Psi(x_{t+1},q_{i})$ for the response. 

Estimating the mutual information from samples is a whole topic in
itself. Recall that the mutual information between two discrete random
variables $X,Y$ is given by \citep{CoverThomas}:

\begin{equation}
I(X,Y)=\sum_{X,Y}p(X,Y)\log\frac{p(X,Y)}{p(X)p(Y)}\label{eq:mutual_inf}\end{equation}

One way to obtain an estimate of the mutual information between two
continuous variables, or between one continuous and one discrete variable
as is the case here, is to estimate their density by computing a histogram
or using a kernel density estimator \citep{ElementsSL}. An estimate
$\tilde{I}$ is obtained by plugging the numbers into equation \ref{eq:mutual_inf},
see \ref{sub:Nonparametric-density-estimation} in the Appendix. 

It turns out that here a faster method is available. Empirically,
we observe that under the Laplace approximation $p(\tau|r_{t+1}=0)$
and $p(\tau|r_{t+1}=1)$ are often well approximated by Gaussians,
especially when the performance wanted is around the mid-point (for
example, 75\% for a 2AFC experiment)%
\footnote{The reason for that is that $\tau_{75\%}\approx\mu$, and $\mu$ has
a Gaussian distribution under the Laplace approximation.%
}. Just as in equation \ref{eq:MIresptheta}, we write:

\[
I(\tau,r_{t+1})=H(\tau)-H(\tau|r_{t+1})\]

Only the latter part of the right-hand side depends on the stimulus
level through $r_{t+1}$. Therefore, to maximise the mutual information,
we minimise :

\begin{equation}
H(\tau|r_{t+1})=-p(r_{t+1}=1)\int p(\tau|r_{t+1}=1)\log p(\tau|r_{t+1}=1)d\tau-p(r_{t+1}=0)\int p(\tau|r_{t+1}=0)\log p(\tau|r_{t+1}=0)d\tau\label{eq:cf_thresholds}\end{equation}

If we model the continous probability distribution functions in this
expression as Gaussians, all we need to do is to compute the mean
and variance of the samples from $p(\tau|r_{t+1}=0)$ and $p(\tau|r_{t+1}=1)$,
the mutual information can then be computed by plugging the numbers
into the expression for the entropy of a Gaussian, then into \ref{eq:cf_thresholds}.
Here again the time complexity of calculating the cost function at
a single point is $\mathcal{0}(n)$, where $n$ is the number of samples.
The number of samples required for a nearly-smooth approximation to
the cost function is however much higher (e.g., 15,000), but in practice
5,000 samples yields good performance if a robust form of optimisation
is used (section \ref{sub:Optimising-T}).

\subsection{Adaptive estimation of width alone}

The $\alpha$-width of a psychometric function is defined for the
Yes/No case in \citet{KussBayesianInferencePsychFunc} as: 

\[
w_{\alpha}=\Psi^{-1}(1-\alpha)-\Psi^{-1}(\alpha)\]

When $\alpha=10\%$, the width is the difference between the 90\%
and the 10\% thresholds. This quantity is useful in describing the
range of a psychometric function - contrary to slope measures, it
is expressed in the same units as the stimulus. In NAFC experiments,
an equivalent measure can be defined by:

\[
w_{\alpha}=\Psi^{-1}(1-\alpha)-\Psi^{-1}(\gamma+\alpha)\]

For a 2AFC experiment ($\gamma=50\%$), and $\alpha=10\%$, $w_{10\%}$
is the difference between the 90\% and the 60\% thresholds.

To compute the mutual information between $w_{\alpha}$ and $r$,
we proceed exactly as with thresholds, with the following modification:
since $w_{\alpha}$ is bounded below by 0, the posterior distribution
is assymetrical. A simple log-transformation renders it symmetrical,
and we can apply our Gaussian method outlined above, or again use
nonparametric density estimation.

\subsection{Adaptive estimation of slope alone}

For a cumulative Gaussian psychometric function such as we the one
we use here the slope is inversely related to $\sigma$; therefore
an appropriate measure of slope is

\[
s=-\log\sigma=-\nu\]

Here again the same technique used for thresholds can be applied:
a Gaussian approximation will do a very good job, but nonparametric
estimation can be used as well.

\subsection{Generalising to any function of the parameters}

The threshold, slope and width as defined above are functions of the
parameters $\theta=(\mu,\sigma,\lambda)$ into the real line. $\tau=\Psi^{-1}(\theta)$,
$w_{\alpha}=\Psi^{-1}(1-\alpha)-\Psi^{-1}(\gamma+\alpha)$ and $s=-\log\sigma$.
It is possible to generalise the algorithm to any function $f(\theta)$
into the real line, with $f$ defined according to the needs of the
experiment. To maximise $I\left(r,f(\theta)\right)$ the algorithm
given for thresholds applies directly, but care should be taken to
check that the distributions $p\left(f(\theta)|r=0\right)$ and $p\left(f(\theta)|r=1\right)$
are well approximated by Gaussians. If not, then an appropriate transformation
or non-parametric estimation should be used, see \ref{sub:Nonparametric-density-estimation}.
Algorithm \ref{alg:The-T-method} presents the T method in its general
form.

\begin{algorithm}
\caption{\label{alg:The-T-method}The T method}

At each time step \emph{t, }for a function $f(\theta)$ defining the
quantity of interest. 

\begin{enumerate}
\item Find the maximum of the log posterior $p_{t}(\theta|\mathbf{y})$
\item Get samples $\mathbf{q}_{\theta}$ from the Laplace approximation
to the log-posterior $p_{t}(\theta|\mathbf{y})$, 
\item For $k$ discrete levels of $x$: compute $\mathbf{q}_{f}=f(\mathbf{q}_{\theta})$,
$\mathbf{q}_{r}=\Psi(x_{i};\mathbf{q}_{\theta})$, and sort the $\mathbf{q}_{f}$
according to $\mathbf{q}_{r}$ to obtain samples from $p_{t}\left(f(\theta)|r\right)$.
Estimate $I\left(f(\theta)||r\right)$ using a parametric or non-parametric
method. 
\item Pick the point $x_{i}$ with the maximal estimated mutual information.
\item Obtain observer response at stimulus level $x_{t+1}$.
\end{enumerate}

\end{algorithm}

\subsection{\label{sub:Optimising-T}Optimising the cost function}

Because it relies on random simulations, the $T$ cost function is
noisy, unlike its $\Psi$ counterpart. Using a plain descent algorithm
in this case is inappropriate, because the derivative information
will be unreliable. A number of measures can be taken to cope with
that problem. We highlight two. 

The simplest is evaluation over a discrete grid, with an optional
iterative refinement step around the minimum over the grid. One find
the minimum over the grid, then expands another grid around the minimum,
and iterates. Depending on the desired precision, the method may or
may not be appropriate. If the possible stimulus levels are constrained
to a discrete set in the first place (for example, contrast on a monitor),
then simply evaluating the cost at every reasonable level makes sense. 

Another option is to first evaluate the cost function over a discrete
grid as above, to fit a smoothing model to the resulting points, and
to optimise over the fitted model. A large number of techniques are
available, including spline models\citep{ElementsSL} and Gaussian
Processes \citep{RasmussenGP}. A minor advantage of splines is that
the minimum between each pair of knots can be computed analytically
(and cheaply), although the primary cost is in the fitting. Interestingly,
the optimal placement of points to find the minimum of a noisy function
is itself an adaptive estimation problem that can be tackled via probabilistic
models\citep{GPnoisyoptimisation}. 

\subsection{Statistical issues}

\subsubsection{Parameterisation, and choice of priors\label{sub:Parameterisation}}

We want to ensure that the posterior distribution is well-approximated
by a Gaussian so that the Laplace method works well. This is difficult
in the $\theta=(\mu,\sigma,\lambda)$ parameterisation because $\sigma\in]0,\infty[$
and $\lambda\in[0,1]$, and the parameter space is therefore bounded.
We follow the recommendations in \citep{GelmanBDA} and use a reparameterisation,
namely $\theta=(\mu,\nu,\eta)$, where:

\[
\nu=\log(\sigma)\]

\[
\eta=\textrm{logit}(\lambda)=\log\frac{\lambda}{(1-\lambda)}\]

This leaves us with parameters defined over $\mathcal{R}^{3}$, which
is perfect for the Laplace approximation.

The natural choice of prior in that space is that of independent Gaussians,
which can be made suitably vague by varying the standard deviation
along each dimension:

\[
p(\theta)=\mathcal{N}\left([m_{\mu},m_{\nu},m_{\eta}],[s_{\mu},s_{\nu},s_{\eta}]\mathbf{I}\right)\]

For $\eta$ fixed the the model is a generalised linear model, and
in the limit $\eta\rightarrow-\inf$ the likelihood $p(\mathbf{y}|\theta)$
is a probit regression likelihood, and guaranteed to be log-concave
\citep{Paninsky-Logconcave}. In practice, to ensure that the posterior
is uni-modal, the prior over $\eta$ should be made precise (ie, lapse-rates
over 4 or 5\% should be very improbable). That is not a major limitation,
since, as Wichmann \& Hill \citeyearpar{Wichmann2001A,Wichmann2001B}
note, lapse rates above 5\% make the measurement of psychometric functions
very difficult and, should they occur, the experiment ought to be
redesigned.

\begin{figure}
\includegraphics[clip,width=12cm]{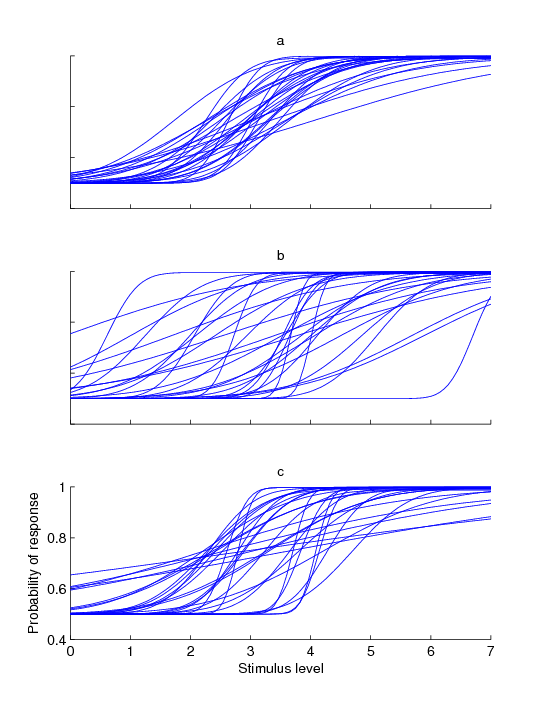}\caption{Adjusting the prior distribution. We display 30 psychometric functions,
drawn from a Gaussian prior distribution. a) Tight prior over both
$\mu$ and $\nu$ b) Looser prior over $\mu$ c) Looser prior over
$\nu$ \label{fig:prior-samples}}

\end{figure}

To get a {}``feel'' for the prior, it is useful to display some
draws from it, as in figure \ref{fig:prior-samples}. The hyper-parameters
can be adjusted until the researcher is confident that the prior contains
realistic psychometric functions.%

\subsubsection{Posterior uncertainty}

\paragraph{Confidence intervals over parameters}

\begin{figure}
\includegraphics[clip,scale=0.7]{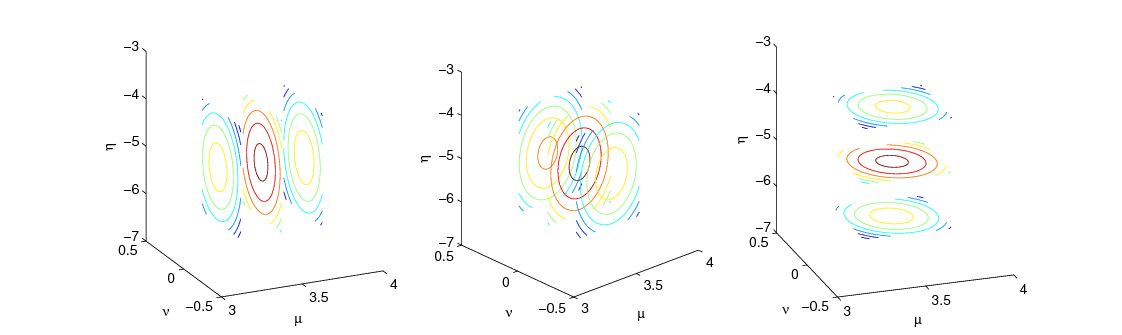}

\caption{\label{fig:pdf-slices}The posterior over parameters in a simulated
experiment, after 200 trials. Slices across the three dimensions reveal
nearly elliptical contours, showing that the Laplace approximation
can be effective. The only aspect that is not well captured by a Laplace
approximation is the slight sharpening of the contours in the direction
of lower $\eta$ (the less an observer lapses, the more informative
his or her responses are). }

\end{figure}

Approximate confidence intervals for $(\mu,\nu,\eta)$ or $(\mu,\sigma,\lambda)$
can be obtained from the empirical quantiles of the posterior distribution,
or from the Hessian at the mode. Confidence intervals over thresholds
$\mathbf{\tau}$ can be obtained in a similar way%
\footnote{Please note that these are Bayesian and not classical confidence intervals,
and can be directly interpreted along the lines of {}``the probability
that the 75\% threshold is between .1 and .2 is 95\%, if my model
is correct''. Should the need to have classical confidence interval
arise, we recommend bootstrapping the maximum likelihood estimates.%
}. The posterior often shows strong correlations between parameters,
so in certain cases it might be worth taking a direct look at the
probability volume, which can be done for instance via a slice representation.
The same type of display can be used to make sure that the Laplace
approximation is a good one (figure \ref{fig:pdf-slices}).

\paragraph{Posterior response distribution}

As a useful display of posterior uncertainty, we also recommend plotting
the posterior response distribution \begin{equation}
p\left(\Psi(x)=q|\mathbf{y}\right)=p(\Psi^{-1}(q;x)|\mathbf{y})\label{eq:postexpresp}\end{equation}
where $\Psi^{-1}(q;x)$ is the $\theta$ such that $\Psi(x;\theta)=q$.
In words, that distribution expresses {}``given the data \textbf{y},
what is the probability that the probability correct of the observer
at stimulus level $x$ is $q$?''.

We plot the quantiles of that distribution for a simulated experiment
in figure X, where color corresponds to quantile rank. Another useful
visualisation is to display draws from the posterior, which can be
compared with draws from the prior (figure X). 

\begin{figure}
\includegraphics[scale=0.6]{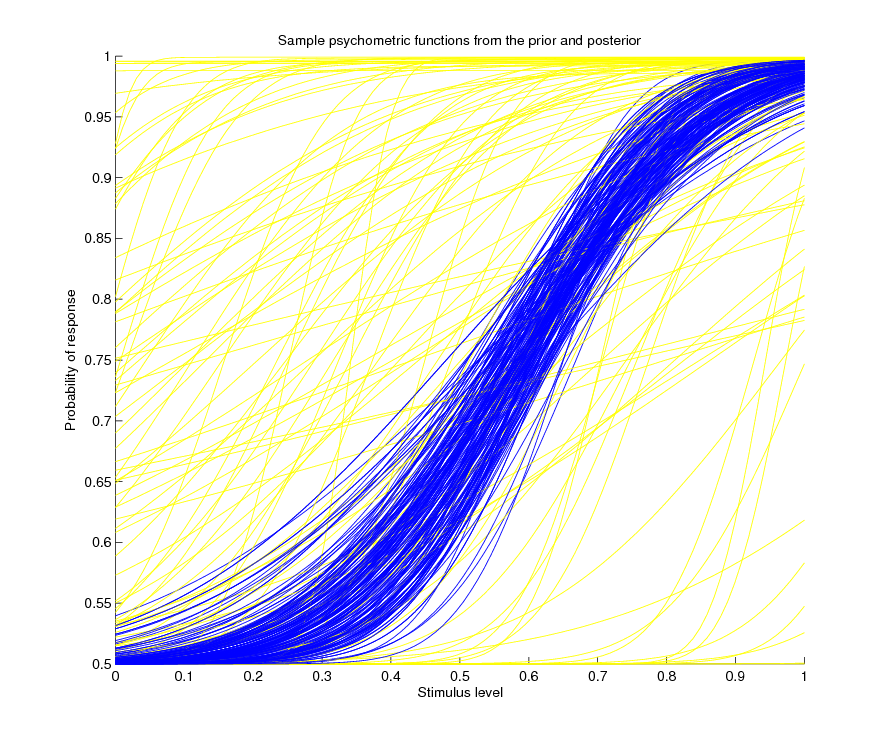}

\caption{\label{fig:posterior-samples}Comparing posterior to prior uncertainty.
After a simulated experiment of 300 trials, samples from the posterior
distribution (in blue) are plotted along with samples from the prior
(in yellow).}

\end{figure}

\begin{figure}
\includegraphics[scale=0.6]{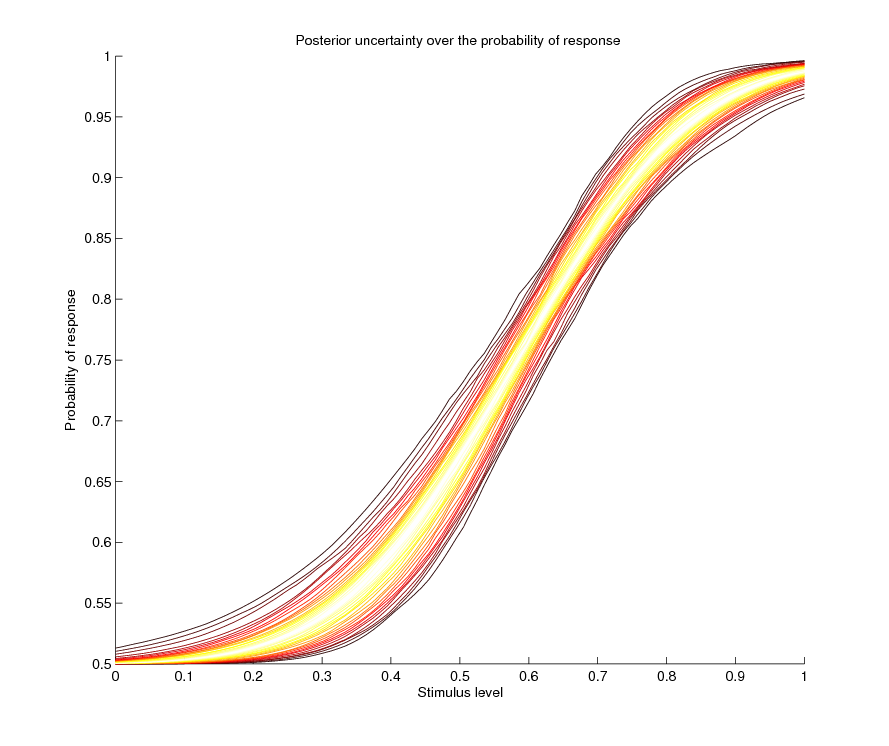}

\caption{Posterior response uncertainty. With the same data as in figure (\ref{fig:posterior-samples}),
we plot the level lines of $p\left(\Psi(x)=q|\mathbf{y}\right)$ with
line color corresponding to quantiles.}

\end{figure}

\subsubsection{Stopping rules}

Instead of stopping the procedure after a pre-set number of steps,
the researcher may want a certain level of confidence to be reached.
A simple solution is to interrupt the procedure after a certain entropy
has been reached. The entropy can be estimated by the entropy of the
Laplace approximation: recall that the covariance matrix is given
by the inverse Hessian at the mode. We see from \ref{eq:entG} that
a stopping rule can be established by setting a minimum level for
$\det(\mathbf{G}^{-1})$, where $\mathbf{G}$ is the Hessian at the
mode.

A more intuitive class of stopping rules is {}``don't stop until
we're X\% sure that a parameter is within certain limits''. This
can be simply read off posterior samples. For example if the rule
is: {}``don't stop until we're 95\% sure that the threshold is above
1'', all one needs to do is to find out the proportion of posterior
threshold samples that lie above 1.

\subsubsection{Estimating response probability}

The usual custom in psychophysics is to fit a psychometric function
to the data, and to use the fitted values for predictions. If the
probability of response at a given stimulus level is of interest then
that is not the best strategy. To estimate that probability of response
it makes more sense to average over values of the parameters: 

\[
p(r;x)=\int_{\theta}p(\theta|\mathbf{y})\Psi(x;\theta)d\theta\]

rather than use the estimated $\hat{\theta}$and compute $\hat{p}(r;x)=\Psi(x;\hat{\theta})$.
Given large amounts of data the difference will be insignificant,
but given small amounts it might not.

\subsubsection{The estimates are biased!}

Correct, if bias is understood in the statistical sense, ie for parameter
$\theta$ and estimator $\hat{\theta}$ as $E(\hat{\theta}-\theta)$.
If you have a prior that says that $\theta$ is more likely to be,
say, 1, while the real theta is .8, then the estimator used here,
the mean of the posterior distribution is going to be some compromise
between what the data are saying (.8), and what the prior is saying.
On the other hand, given the fact that $\theta$ is unknown, and whatever
we know of it is contained in the prior and in the data, the mean
of the posterior represents our best guess - choosing an unbiased
estimator would imply disregarding the prior information, leading
to far worse performance in practice.

\subsubsection{How do I know my prior is correct? Can I avoid using one? }

There are different aspects to that question. The first one is the
question of how the use of a prior influences the final estimates.
If the data are sufficient, and the prior is not too precise, the
answer is not much at all. It is very easy to check for this: once
the data have been collected, simply re-estimate the parameters of
interest with a loosened prior. In the limit of high prior variance,
the maximum a posteriori estimate tends to the maximum likelihood
estimate, and in the limit of large data as well. When the data do
not constrain the parameters very well, the maximum a posteriori estimate
and the maximum likelihood estimate may be quite different. In that
case, the maximum likelihood estimate can be clearly non-sensical,
and prior constraints will help getting something sensible. What to
do depends on what the data are needed for: if the adaptive procedure
is used prior to an experiment to set levels appropriate to an observer's
performance, then good prior information can compensate to some extent
for not-so-good data. If the adaptive procedure is used to collect
the main data, then estimates that are not very robust to prior assumptions
may be really bad news. Devising clever statistical procedures based
on hierarchical models \citep{GelmanBDA} is a way of salvaging such
data but running more subjects might be the simplest solution. 

\begin{figure}
\includegraphics{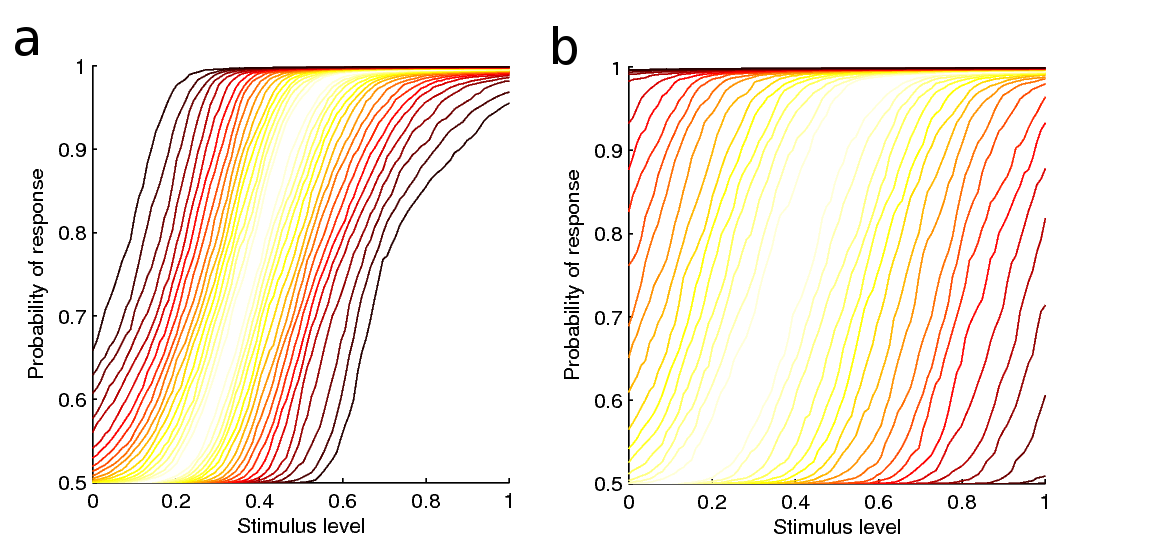}

\caption{\label{fig:Adjusting-the-prior}Adjusting the prior using the prior
response probability. The level curves in the above figures are level
lines of $p(\Psi(x))=\Psi(x;\theta)p(\theta)$ for a prior that is
tight in $\mu$ (figure a) and one that is very wide (figure b). In
the context of a detection experiment, suppose a stimulus level of
1 is the highest luminance that can be displayed. Clearly if the observer
is still at chance for that level there is something wrong with the
experiment or with the observer. Knowing that, the prior response
probability of figure a is more appropriate than that of figure b. }

\end{figure}

The second, potentially separate issue is the use of prior information
for adaptive measurement. Such information is essential to avoid wasting
trials testing observers on stupidly high or low stimulus intensity.
Fortunately, this is quite easy to do. The software package we wrote
includes a graphical utility for setting a prior that makes sense.
As an experimenter, some simple considerations can help decide how
to tune the parameters of the prior. For example, in a detection task,
if an observer cannot detect the stimulus at 100\% contrast, something
must be going really wrong. Plotting prior proabibility of response,
as in figure \ref{fig:Adjusting-the-prior}, helps modify the prior
to take that fact into account. It is also safe to assume that if
the overtrained experimenter can't see the stimulus the subjects won't
be able to either. Again, it's fairly easy to set the prior so that
simple constraint is taken into account. Setting a prior on the slope
is not too difficult either: do we believe observers can go from 50\%
correct to 95\% correct in the space of a contrast increment? If yes,
then prior samples should show that feature, if not then they should
not (see figure \ref{fig:prior-samples}). 

\subsubsection{\label{sub:I-can't-check-the-fits}I can't check the fits by eye!}

When using the method of constant stimuli, psychophysicists like to
plot the measured proportion at each level along with the fit. The
fit is then deemed to be good if it appears to go more or less through
the datapoints.

One caveat is that psychometric functions are usually plotted on a
probability scale because everyone likes a nice S-shape. This is unfortunately
inappropriate for judging a fit because, unlike in least-squares regression,
binomial error can't be assessed from Euclidean distances in a plot
on a probability scale. Predicting 50\% and measuring 60\% is not
bad, predicting 90\% and measuring 80\% can be. The problem can be
alleviated by plotting the psychometric function on a probit scale,
which will make the psychometric function almost linear, and by providing
sampling intervals for the fitted psychometric function. The sampling
intervals answer the question: if the model is true, where would the
central 95\% of the measured values lie? This is given by the binomial
distribution, is asymetrical and therefore does not map to the Euclidean
metric we use to {}``eyeball'' fits.

When running an adaptive experiment, we measure responses at many
different points along the stimulus level axis, and we do it only
once. It is not possible to plot the proportion of responses in this
way, and therefore the familiar plot of the fitted psychometric function
is impossible. Not all is lost, because there are alternatives.

A bad alternative is to bin together responses from nearby levels
(this bad idea has nonetheless been used in several papers we won't
cite). This does not make any sense because we cannot, on the one
hand, assume that the probability of response varies continuously
with the stimulus level and then act as if it did not, by binning.
Furthermore, there is no unique criterion to determine the size of
the bins, and sizes could be hand-picked such as to make the fit look
good or bad. 

A simple alternative is to plot the responses as dashes along the
stimulus level axis, as shown in figure X. This makes it possible
to diagnose really bad fits (arising from a completely mistaken prior,
either too flat or too peaked), and to assess whether the psychometric
function has been properly sampled.

A more sophisticated option is to use posterior predictive checks
\citep{GelmanBDA}. The intuition behind the method is that models
that fit well should give the observed data high probability under
the posterior. The posterior predictive distribution is given by:

\[
p(\mathbf{y^{*}}|\mathbf{y})=\int_{\theta}p(\mathbf{y^{*}}|\theta)p(\theta|\mathbf{y})d\theta\]

It represents the probability of getting data $\mathbf{y}^{*}$ when
we've observed data $\mbox{\textbf{y}}$. It is easy to generate samples
from that distribution given samples from the posterior $p(\theta|\mathbf{y})$
- see \ref{sub:t-method}. Posterior predictive checks compare the
actual, measured data from simulated datasets obtained from the posterior

One possible problem that could arise when estimating a psychometric
function is when performance changes over time, for example due to
drifts in attention levels. An example diagnostic plot is given in
figure X. Here we simulate an observer whose psychometric function
changes in the course of an experiment: the $\mu$ parameter is under
a downward drift, $\mu(t)=3.5$, and the other parameters are set
as in \ref{sub:Psi-method-convergence-speed}. 

\begin{figure}
\includegraphics[clip,scale=0.6]{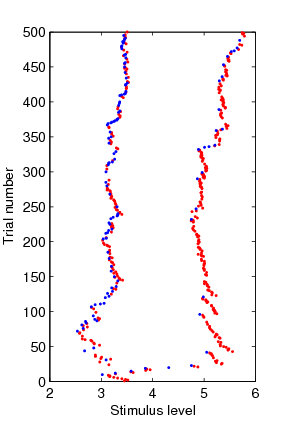}\caption{\label{fig:PPC-real}Posterior predictive checks. We plot the result
of a simulated run of the $\Psi$ algorithm for an observer whose
psychometric function changes over time (see text). Correct responses
are plotted in red, incorrect responses in blue.}

\end{figure}

\begin{figure}
\includegraphics[clip,scale=0.8]{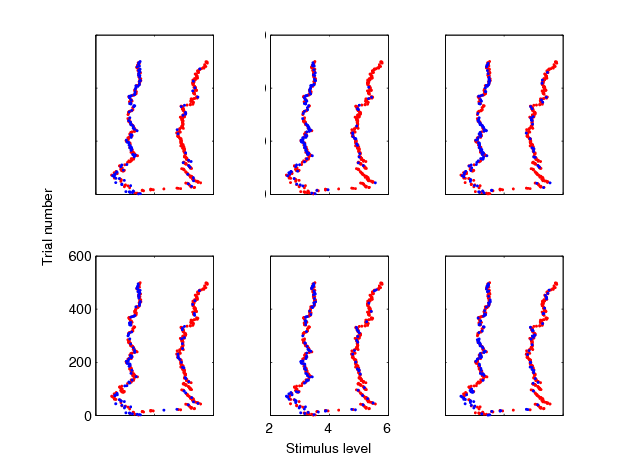}\caption{\label{fig:PPC-simul}Simulated datasets from the posterior predictive
distribution. Discrepancies with the observed data are visible. }

\end{figure}

We plot level, trial, response triplets $(i,x_{i},r_{i})$ with stimulus
level along the x axis, trial number (time) along the y axis, and
the response, correct or incorrect, is color coded. The actual dataset
is plotted in figure \ref{fig:PPC-real}, and in figure \ref{fig:PPC-simul},
we plot simulated datasets from the predictive distribution. The lack
of fit can be seen from the overprediction of correct responses in
the late trials and the underprediction in the early trials.

\section{Simulation results}

We conducted an empirical evaluation of our software based on simulations.
We report results on speed of convergence with respect to other sampling
schemes, robustness to a misspecificed prior, and a comparison with
the QUEST procedure.

\subsection{The $\Psi$ method: speed of convergence\label{sub:Psi-method-convergence-speed}}

How well does the $\Psi$ method perform compared to other sampling
schemes? We tested the following:

\begin{enumerate}
\item Uniform sampling from an interval. In this case the observer is tested
at levels drawn uniformly from an interval. We test varied the spread
of the interval, from wide (between 50.01\% and 98.5\% correct), medium
(between 55\% and 95\% correct) to tight (the interval only covers
values of the psychometric function between 70\% correct and 85\%
correct). 
\item Constant stimulus: we picked 6 stimulus levels for testing the observer.
Here again we varied the spread of the levels: we took the same 3
intervals used for uniform sampling and divided them into 5 parts
of equal length. 
\item Our version of the $\Psi$ method.
\end{enumerate}
We took the case of a 2AFC experiment. The lapse rate of the simulated
observer was set to 2\%. The simulated psychometric function is show
in figure X, along with the sampling schemes we used. We varied the
number of trials from 50 to 500, and took the mean a posteriori of
$\mu$ and $\nu$ given the data as our estimates. We show the m.s.e.
as a function of number of trials for the different conditions in
figure X. We set the {}``true'' parameters to $\mu=3.5$, $\nu=0.5$,
with priors:

\[
\mu\sim\mathcal{N}(3,\sqrt{0.5})\]

\[
\nu\sim\mathcal{N}(0,\sqrt{0.5})\]

\begin{figure}
\includegraphics[clip,scale=0.3]{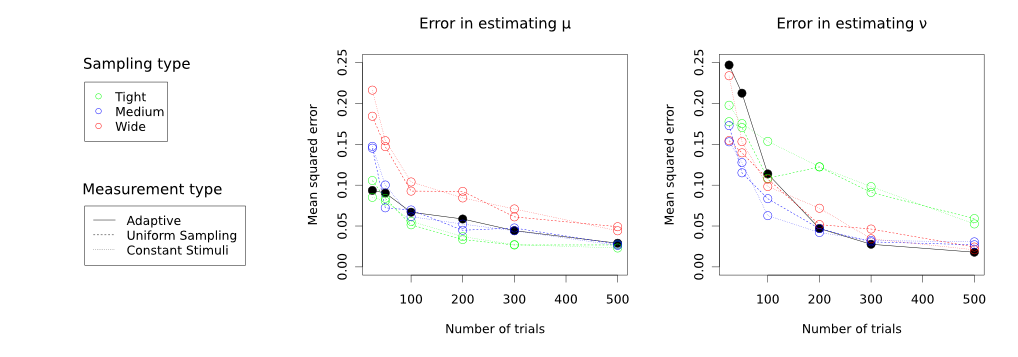}

\caption{Comparison of measurement methods and sampling schemes. Mean squared
error is plotted as a function of number of simulated trials (each
datapoint represents an average over 150 simulated experiments) .
\label{fig:Comparison-psi-vs-cs}}

\end{figure}

The results are shown on figure \ref{fig:Comparison-psi-vs-cs}.

An important observation is that a tight interval around the threshold
yields a good estimate of $\mu$ but a bad estimate of $\nu$, and
vice-versa for a wide estimate. The $\Psi$ method seeks a compromise
between the two, and achieves much better performance than the medium
'trade-off' sampling scheme, as is evident from the figures. 

\subsection{Robustness}

In the paragraph above, the true values of the parameters are each
one standard deviation away from the assumed prior mean. We'd like
to know how the $\Psi$ method behaves when true values turn up to
be very unlikely under the prior, ie. how robust it is to prior assumptions.

We repeat the experiment above, still setting $\mu=3.5$ and $\nu=0.5$,
but starting with the following priors

\begin{figure}
\includegraphics[clip,scale=0.5]{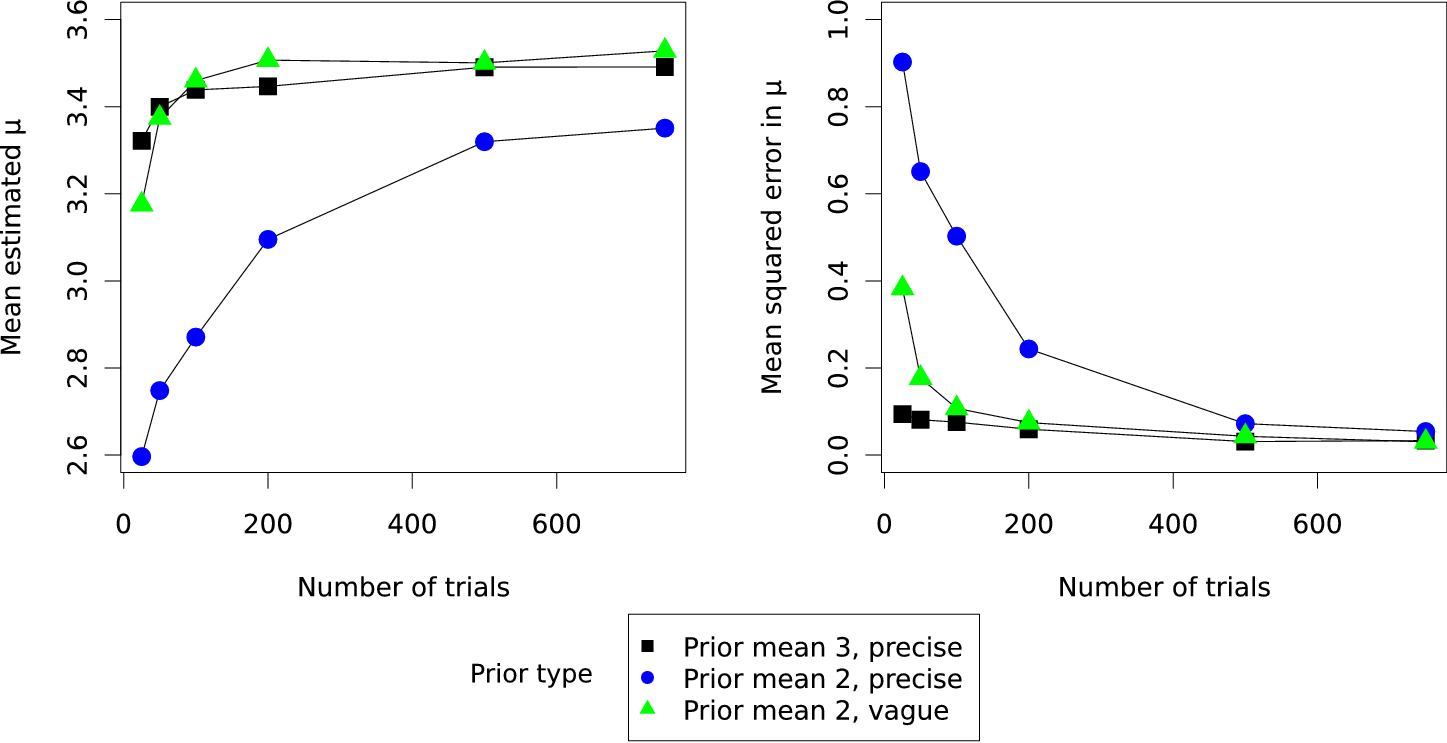}

\caption{\label{fig:Robustness}Experiments were simulated using three different
types of priors. The mean estimate of $\mu$ and the associated mean
square error are displayed as a function of the number of trials.
Each datapoint represents an average over 150 simulated experiments.}

\end{figure}

\begin{enumerate}
\item $\mu\sim\mathcal{N}(3,\sqrt{0.5})$: true value less than one standard
deviation away, precise prior
\item $\mu\sim\mathcal{N}(2,\sqrt{0.5})$ true value more than two standard
deviations away, precise prior
\item $\mu\sim\mathcal{N}(3,1)$ true value 1.5 standard deviations away,
vague prior
\end{enumerate}
The results are plotted on figure \ref{fig:Robustness}. Prior (2)
giving little prior probability to the true value starts off dramatically
bad, but shows roughly exponential recovery. Prior (3) shows less
bias and catches up with prior (2) quite fast. Bayesian asymptotics
guarantee that as long as the prior gives non-null probability to
the true value of the parameter, the maximum of the posterior will
converge to the true value in the limit of large data. The lesson
here is that vague priors are the prudent way to go if the circumstances
allow. One way to check for large violations of prior assumptions
during data analysis is to compare MAP and ML estimates to make sure
there are no systematic discrepancies. 

\subsection{T method versus QUEST}

Although now a quarter of a century old, the QUEST algorithm remains
a widely-used tool for the adaptive measurement of psychometric thresholds,
one of the reasons being that it is available in a quality implementation
in the popular package PsychToolbox \citep{Brainard:1997,Pelli:1997}.
In this simulation we compare QUEST's performance to that of the T
method. 

For the comparison to be fair, three issues had to be settled. The
first issue is that QUEST assumes that the psychometric function has
Weibull shape, whereas we assume it is a cumulated Gaussian. We account
for that by simulating alternatively an observer with Weibull shape
and an observer with Gaussian shape. We set the parameters for the
Gaussian observer to $\mu=6,\nu=0.5$, and adjusted the parameters
$\alpha,\beta$ of the Weibull observer so that the shape of the psychometric
function was as close as possible, as measured by $L_{2}$ distance.

\begin{equation}
argmin_{\alpha,\beta}\int_{0}^{+\inf}\left(\Psi(x;\mu,\nu,\eta)-\Psi_{w}(x;\alpha,\beta,\eta)\right)^{2}dx\label{eq:minL2}\end{equation}

\begin{figure}
\includegraphics[scale=0.5]{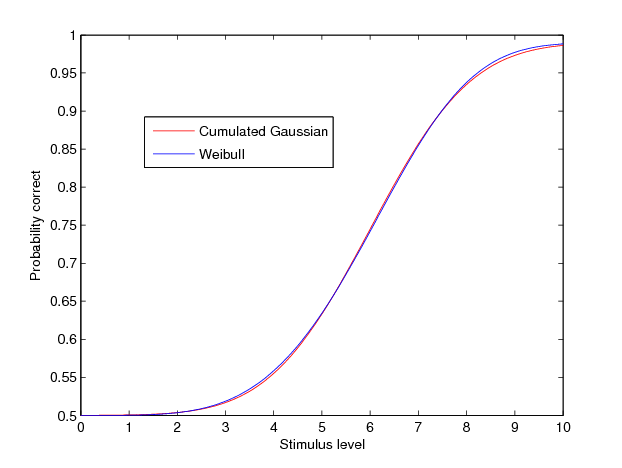}

\caption{The psychometric function of the simulated Weibull observer was adjusted
so as to match that of the cumulated Gaussian observer as closely
as possible, as measured by $L_{2}$ distance. The resulting psychometric
function is plotted in blue, with the target in red.\label{fig:Weibull-vs-Gaussian}}

\end{figure}

The Weibull psychometric function is a variant of \ref{eq:psychfuncFC}:
\begin{equation}
\Psi_{w}(x;\mu,\sigma,\lambda)=(1-\lambda)\left(\gamma+(1-\gamma)w(x;\alpha,\beta)\right)+\lambda\gamma\label{eq:psychfunc_Weibull}\end{equation}
$w(x;\alpha,\beta)=1-\exp\left(-\left(\frac{x}{\alpha}\right)^{\beta}\right)$
is the Weibull function. The integral \ref{eq:minL2} has no analytical
form, so we used numerical integration and numerical optimisation.
The result of the adjusment is displayed on figure \ref{fig:Weibull-vs-Gaussian}.
To keep things even half of the trials were run with the Weibull observer,
and half without. 

\begin{figure}
\includegraphics[scale=0.5]{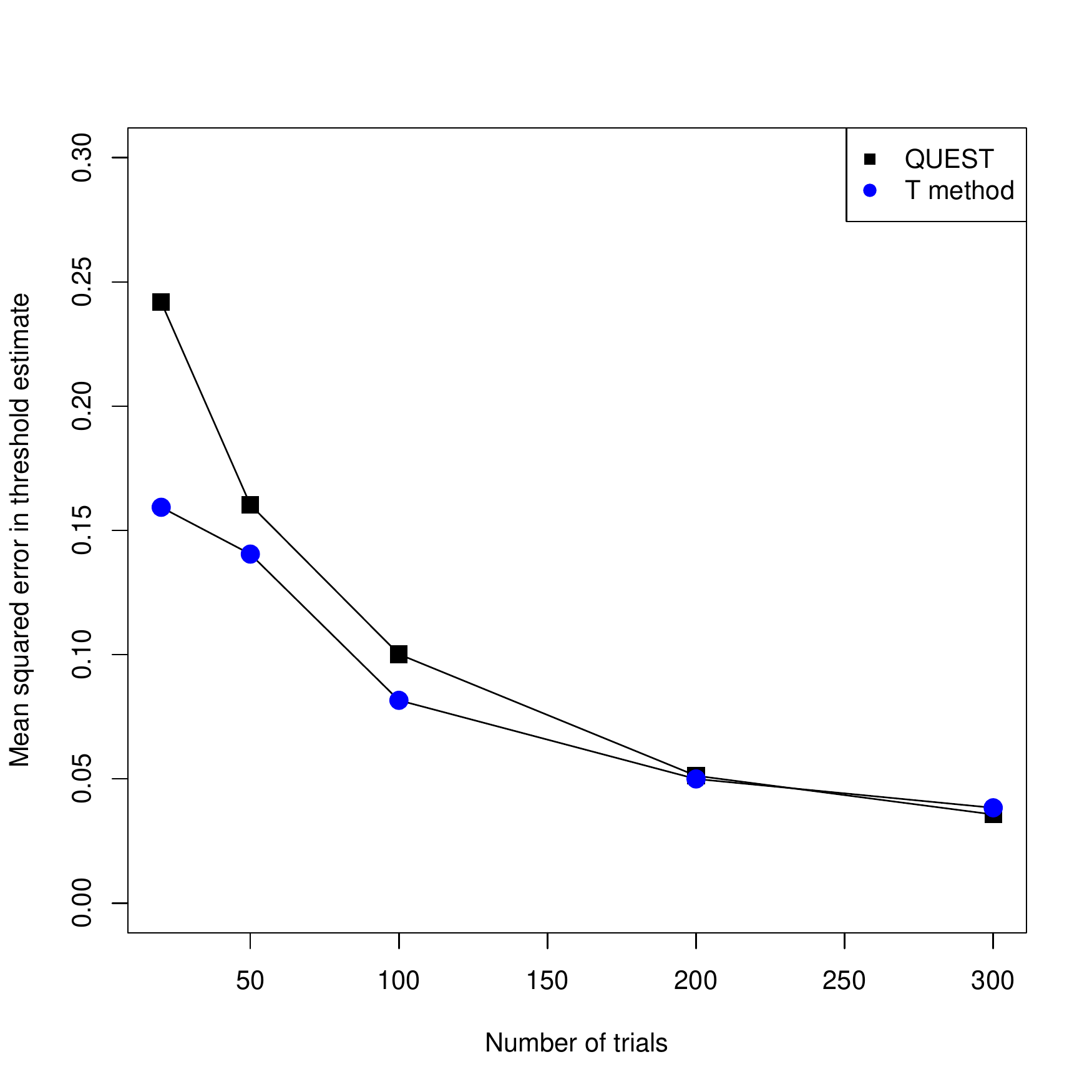}

\caption{Comparison of QUEST and the T method. We plot the mean squared error
in estimation of the 75\% threshold (each datapoint represents an
average over 100 simulated experiments).\label{fig:QUEST-vs-T}}

\end{figure}

The second set of issues has to do with the prior assumptions embedded
in the QUEST algorithms. We needed to ensure that the comparison was
fair in terms of assumed prior knowledge, that QUEST did not know
more \emph{a priori }about the parameters than our method and vice-versa: 

\begin{itemize}
\item QUEST asks the user to provide a {}``guess'' for the threshold along
with a prior variance. We therefore initialised QUEST with a guess
and prior variance for the threshold corresponding to the mean and
variance of $p(\tau|\mu,\nu,\eta)$, obtained by simulating from the
prior for the T method and computing the threshold for each sample
(as in section \ref{par:A-posteriori-probability-over-thresh}). 
\item QUEST assumes a known parameter$\beta$ in \ref{eq:psychfunc_Weibull},
but this parameter is never known exactly in practice. So QUEST's
$\beta$ value was set each time to a different, random, value to
express the prior uncertainty in this parameter. Again, we sampled
from the prior $p(\mu,\nu,\eta)$ and computed the best fitting $\alpha,\beta$
values by optimisation of \ref{eq:minL2} as explained above. $p(\log\beta|\mu,\nu,\eta)$
had a Gaussian distribution with mean 2.33 and standard deviation
0.77. Therefore, in every simulated QUEST experiment, we set $\log\beta$
to a random value from $\mathcal{N}(2.33,0.77)$.
\item QUEST's performance depends on a precision parameter. We increased
the precision parameter until no further visible performance gain
could be obtained. The performance of the T method can be increased
by using more samples from the posterior, by refining the optimisation
grid, by using non-parametric estimation, etc. We set the number of
samples to 5000, the number of grid points to 45, and did not use
non-parametric estimation. Those settings keep computing time to a
reasonable 280ms per trial on the not-so-recent computer we used to
run the tests. 
\end{itemize}
We used the functions QuestQuantile to determine stimulus level and
QuestMean for the final estimate. The results are shown on figure
\ref{fig:QUEST-vs-T}. The T method has a lead when the number of
trials is low, but at 200 trials the methods yield essentially identical
results. Although some marginal improvements could still be had in
the T method by increasing the computational load, this leads us to
think that in the case of estimating thresholds we are close to ceiling
performance, unless we move to more sophisticated models than the
psychometric function (e.g., models that take into account possible
interdependencies between trial outcomes instead of assuming independence). 

\section{Conclusion}

Our framework can easily be extended to a range of related problems,
such as the adaptive measurement of psychometric surfaces (probability
of response as a function of more than one stimulus dimension), as
in \citet{KujalaLukka06}. A more challenging extension and a direction
for future work is to model-based adaptive measurement in psychophysics
\citep{QuickTvC}. The usual course of action in psychophysical experiments
is to acquire some data first, and then to fit models to evaluate
hypotheses. Using the techniques outlined here it is possible to select
stimuli to minimise the uncertainty about model parameters, or even
to select stimuli that help determine which model is more appropriate
- an adaptive form of hypothesis testing. We are currently investigating
an application to the Bayesian estimation of classification images.

\section{Appendix}

\subsection{Bayesian inference for psychometric functions}

\subsubsection{Computing the posterior}

The posterior over the parameters $\theta=(\mu,\nu,\eta)$ given the
data is given by:

\[
p(\theta|\mathbf{y})=\frac{p(\mathbf{y}|\theta)p(\theta)}{p(\mathbf{y})}\]

Where the likelihood is equal to:

\[
p(\mathbf{y}|\theta)=p\left(\{\mathbf{r},\mathbf{x}\}|\theta\right)=\prod_{i=1}^{N}\left(r_{i}\Psi(x_{i},\theta)+(1-r_{i})(1-\Psi(x_{i},\theta))\right)\]

The posterior has no closed form solution, so an approximation must
be used. \citet{KussBayesianInferencePsychFunc} use reversible-jump
Markov Chain Monte Carlo, but we found that the much simpler Laplace
approximation works well enough for most purposes, given the appropriate
parameterisation (section \ref{sub:Parameterisation}). 

If something more precise is needed, the posterior is only three-dimensional
so traditional numerical integration, while slow, is still tractable.
We recommend integrating over a cubic region that includes most of
the mass of the Laplace approximation, in case the Laplace approximation
is much sharper than the actual density. The density is continuous
and twice differentiable so traditional integration rules apply.

\subsubsection{Integrating out the lapse rate}

In many cases the lapse parameter $\eta=\textrm{logit}(\lambda)$
is a nuisance variable of no direct interest. The standard way of
dealing with nuisance variables in Bayesian statistics is by {}``integrating
out'':

\begin{equation}
p(\mu,\nu|\mathbf{y})=\int_{-\inf}^{+\inf}p(\mu,\nu,\eta|\mathbf{y})d\eta\label{eq:integrating-out}\end{equation}

Under the Laplace approximation, $p(\mu,\nu,\eta|\mathbf{y})$ is
multivariate Gaussian and therefore so is $p(\mu,\nu|\mathbf{y})$,
and the correlation matrix is unchanged \citep{RasmussenGP}. When
working with samples from the posterior if $(\mu^{*},\nu^{*},\eta^{*})$
are samples from $p(\mu,\nu,\eta|\mathbf{y})$ then $(\mu^{*},\nu^{*})$
are samples from $p(\mu,\nu|\mathbf{y})$.

\subsection{The psychometric function in Yes/No experiments}

In Yes/No experiments the range of the psychometric function is $[0,1]$.
We adopt the following functional form:

\begin{equation}
\Psi(x;\mu,\sigma,\lambda)=(1-\lambda)\Phi(x;\mu,\sigma)+\frac{\lambda}{2}\label{eq:psychfunc-yn}\end{equation}

The parameters have the same interpretation as in the forced-choice
case, see \ref{sub:Functional-form}.

\subsection{\label{sub:Reduction-mult-thresh}Reduction of the multiple-threshold
case to the $\Psi$ method}

Suppose that the researcher wishes to find appropriate stimulus levels
to put the observer at 0.65, 0.75 and 0.85 performance. In other words,
he or she needs to estimate a vector $\mathbf{\tau}$ of thresholds.
In cases maximising the mutual information between $\mathbf{\tau}$
and the observer's responses is equivalent to maximising the mutual
information between $\theta$ and the observer's responses (approximately
for two thresholds, exactly for three and more). In other words, the
$\Psi$ method is appropriate in this case. 

This is quite easy to see: if we know $\theta$ then $\tau$ is known
(the psychometric function defines unique thresholds). On the other
hand, if we know that $\Psi(x_{1};\mu,\sigma,\lambda)=q_{1}$,$\Psi(x_{2};\mu,\sigma,\lambda)=q_{2}$,$\Psi(x_{3};\mu,\sigma,\lambda)=q_{3}$
(ie we know three thresholds), then we know $\theta=(\mu,\sigma,\lambda)$\emph{.
}Each known threshold fixes one parameter with respect to the others.
Therefore there exists a one-to-one correspondence between $\theta$
and $\mathbf{\tau}$, so that in probabilistic terms, a density over
$\theta$ is a transformation of a density over $\tau$, and vice
versa. The mutual information is invariant to transformations of the
variables, therefore $I(\mathbf{\theta},\mathbf{r})=I(\mathbf{\tau},\mathbf{r})$
and the $\Psi$ method applies.

In the case where $\tau$ is two-dimensional then the one-to-one correspondence
does not exist, and: 

\[
p\left(\tau_{q_{1},q_{2}}=(x_{1},x_{2})\right)=\int_{\Theta}I_{\theta,\tau}p(\theta)d\theta\]

where $I_{\theta,\tau}$ is the indicator function of the solution
set of $\left(\begin{array}{cc}
\Psi(x_{1};\theta) & =q_{1}\\
\Psi(x_{2};\theta) & =q_{2}\end{array}\right)$, which contains all the $(\mu,\sigma,\lambda)$ triplets compatible
with $q_{1},q_{2}$ thresholds being at $x_{1},x_{2}$. We'll make
a heuristic argument of why in this case the $\Psi$ method is still
a good approximation. 

Notice that $\lambda$, the lapse rate, affects thresholds very little
within its normal range ($\lambda$ < 5\%), especially for thresholds
far from the extremes. So most of the uncertainty about the thresholds
can be traced to uncertainty about $\mu$ and $\sigma$. Consequently,
given two thresholds $\tau_{1}$ and $\tau_{2}$ known exactly, there
is little uncertainty left in the estimation of $\mu$ and $\sigma$.
This implies that having knowledge of $\theta$ and having knowledge
of $\tau=(\tau_{1},\tau_{2})$ is almost the same thing, and gaining
information about one is almost the same as gaining information about
the other.

\subsection{Nonparametric density estimation for the T method\label{sub:Nonparametric-density-estimation}}

The problem of estimating the mutual information between threshold
(or width, or slope) and response hinges on the computation of the
conditional densities of $p(\tau,r)$: $p(\tau|r)$, or, equivalently,
$p(r|\tau)$. The first (a continuous density as a function of a binary
variable) can be dealt with using non-parametric density estimation
techniques, the second (a probability distribution over a binary variable
as a function of a vector of continuous variables) can be dealt with
using a multiplicity of techniques developed for logistic regression.

\subsubsection{Kernel density estimation }

Kernel density estimation is a natural extension of the well-honed
technique of building histograms \citep{ElementsSL}. A histogram
estimates a density $f_{X}$ by binning the data $\mathbf{X}=(\mathbf{x_{1},}\mathbf{x_{2},...,}\mathbf{x_{N})}$: 

\[
\hat{f_{X}}(x)=\frac{1}{N}\times B(x;\mathbf{X})\]

The function $B(x,\mathbf{X})=\sum_{i=1}^{N}I_{x}(x_{i})$, where
$I_{x}$ is the indicator function for the bin that contains $x$,
simply counts how many samples are in the same bin. Within that neighbourhood,
all samples are given equal weighting, and without, all are discarded.
This leads to non-smooth density estimates (as we've all noticed,
histograms are blocky).

In kernel density estimation, to estimate $f_{X}(x)$, we weigh samples
according to how far they are from $x$, with the weighting given
by a kernel function. The kernel density estimate is given by \[
\hat{f}_{X,t}(x)=\frac{1}{Nt}\sum_{i=1}^{N}K(x,x_{i})\]

$K(x,x')$ is a kernel function, generally a Gaussian (we promise
this is the last occurence of the word Gaussian in this article).
The smoothness parameter $t$ is estimated from the data. The kernel
density estimator has better convergence properties than the histogram
for smooth distributions. Once we have estimated $t$, the entropy
estimate for the density $f_{X}(x)$ follows naturally

\[
\hat{H}(f_{X})=-\int\hat{f}_{X,t}(x)\log\hat{f}_{X,t}(x)\]

This quantity can be computed numerically quite easily.

\subsubsection{Advantages and shortcomings}

The main advantage of the kernel density estimator is that it is non-parametric.
The main problem here is that density estimation or non-parametric
logistic regression occurs within an optimisation loop, so that time
complexity is the major issue. If we require \emph{m }evaluations
of $\hat{f}_{X,t}$ to compute the entropy estimate based on \emph{n}
samples, a naive implementation would have $\mathcal{O}(mn)$ time
complexity. A clever implementation based on the Fast Fourier Transform
can bring that down to $\mathcal{O}(m\log n)$.

As a proof of concept, we have implemented support for Gray and Moore's
\citep{GrayTreeDensityEstimation} kernel density estimator, which
uses a tree structure to speed up processing. Alternatively, kernel
logistic regression is a candidate for modelling $p(r|\tau)$. Unfortunately,
the computational cost is still severe enough to make it prohibitive
on all but the most recent desktop computers. Future advances will
no doubt make this practical.

\bibliographystyle{apalike2}
\bibliography{15_home_simon_Documents_Papers_BayesAdapt_ref}

\end{document}